\documentclass[letterpaper,12pt]{article}
\usepackage{tabularx} 
\usepackage{amsmath}  
\usepackage{graphicx} 
\usepackage[margin=1in,letterpaper]{geometry} 
\usepackage[final]{hyperref} 
\hypersetup{
	colorlinks=true,       
	linkcolor=blue,        
	citecolor=green,        
	filecolor=magenta,     
	urlcolor=blue
}
\usepackage{graphicx}
\usepackage{enumitem}
\usepackage{color}
\usepackage[affil-it]{authblk}
\usepackage{cite}

\begin{document}
\title{\textbf{A Model of the Collapse and Evaporation of Charged Black Holes}}

\renewcommand*{\Authfont}{\normalsize}  
\author[1]{Yu-Ping Wang }
\affil[1]{Department of Physics, National Taiwan University, Taipei 106, Taiwan, R.O.C. \footnote{Email: \href{mailto:b03202045@ntu.edu.tw}{b03202045@ntu.edu.tw}}}
\date{}
\maketitle
\begin{abstract}
In this paper, a natural generalization of KMY model is proposed for the evaporation of charged black holes. Within the proposed model, the back reaction of Hawking radiation is considered.  More specifically, we consider the equation $G_{\mu\nu} = 8\pi \langle T_{\mu\nu}\rangle$, in which the matter content $\langle T_{\mu\nu}\rangle$ is assumed spherically symmetric. With this equation of motion, the asymptotic behavior of the model is analyzed.  

Two kinds of matter contents are taken into consideration in this paper. In the first case (the thin-shell model), the infalling matter is simulated by a null-like charged sphere collapsing into its center. In the second case, we consider a continuous distribution of spherical symmetric infalling null-like charged matter. It is simulated by taking the continuous limit of many co-centric spheres collapsing into the center.

We find that in the thin-shell case, an event horizon forms and the shell passes through the horizon before becoming extremal, provided that it is not initially near-extremal. In the case with continuous matter distribution, we consider explicitly an extremal center covered by neutral infalling matter, and find that the event horizon also forms. The black hole itself will become near-extremal eventually, leaving possibly a non-electromagnetic energy residue less than the order of $\ell_{p}^{4}/e_{0}^{3}$.

The details of the behavior of these models are explicitly worked out in this paper.
\end{abstract}

\section{Introduction} \paragraph{ }
As to the study of Hawking radiation, it is often assumed that the background geometry is that of the classical black holes, such as Schwarzschild metric for neutral non-rotating black holes, Reissner-Nordström metric for charged non-rotating black holes, etc. The back reaction of Hawking radiation is usually neglected in these cases, assuming that it has little impact on the relevant physical features of the systems.\cite{ hawking1975particle,parker2009quantum, siahaan2010semiclassical, triyanta2013hawking}
%

In the paper of Kawai, Matsuo and Yokokura, a new consistent model of black hole evaporation was proposed.\cite{kawai2013self,kawai2016interior,kawai2017model} This model (KMY model) is a consistent solution to the semi-classical Einstein equation:
\begin{equation}
G_{\mu\nu} = 8\pi \langle T_{\mu\nu}\rangle
\end{equation}
where $\langle T_{\mu\nu}\rangle $ consists of Hawking radiation and the infalling matter. As a special case of KMY model, one single null-like shell of infalling matter is considered in Schwarzschild background.\cite{kawai2013self} It was found out that, in this case, the black hole would never evaporate completely, and event horizon and singularity are formed. In the paper, a more general case with a continuum of infalling null shells was also taken into consideration.\cite{kawai2013self} In this general case, the black hole evaporates completely, and each layers of collapsing matter never touches its corresponding Schwarzschild radius.

This model gives us many useful insights to the problem of information loss paradox\cite{ho2015comment, ho2016absence}, since in KMY model, either a permanent event horizon forms, or the infalling matter never reaches event horizon. It is  unnecessary to introduce any non-local interaction to maintain the unitarity of the initial state of infalling matter and the final state of Hawking radiation.

It is therefore natural to ask whether this model can be generalized to more complex cases, such as charged or rotating black holes. We also hope that the beautiful properties of KMY model, such as that the infalling matter will never pass through event horizon, or the formation of a permanent event horizon, will persist after this generalization. These properties imply that the information will not pass through event horizon, or will be locked in a permanent horizon.

In this paper, we generalized KMY model to describe the evaporation of charged black holes, considering the thin-shell model (a charged null-like collapsing sphere) and the a more general model of continuously collapsing matter (the continuous limit of many co-centered charged null-like shells). A detailed analysis showed that the information of infalling matter will not be lost in either case. During the analysis, we had assumed that Hawking radiation is both chargeless and massless, which is a fairly reasonable assumption if the Hawking temperature is low, since all charged particles are massive in the standard model, and if the Hawking temperature of this field is way lower than the rest mass of the field, then only a little radiation can be emitted. We could also show that, for this model, the Hawking temperature is indeed very low at any time during the evaporation, assuming that it is initially very small.

We also showed in this paper that, an event horizon will always exist when a black hole is charged, whether in the thin-shell case or in the case of continuously collapsing matter. The difference is that, in the thin-shell model, the black hole only emits certain amount of radiation before it enters the horizon, and the black hole remains non-extremal. On the other hand, in a model of continuously collapsing matter, there could be two cases depending on the initial condition. In the first case, all layers of shells become extremal as they pass the horizon. In the second case, there exists a critical layer. A shell above the critical layer will become extremal as it enters the horizon, while a shell below the critical layer will not become extremal as it enters the horizon\footnote{Under this case, there is still a possibility that the critical layer is the outermost layer, and the black hole does not become totally extremal. We will show in this paper that even if this is the case, the residue of the non-electromagnetic energy is at the order of $\frac{\ell_{p}^{4}}{e_{0}^{3}}$.}. One can see this better if we present the black hole in the Penrose diagram (See Fig.\ref{fig:1-2} and Fig.\ref{fig:1-1}).

This paper is presented in the following way. In section \ref{sec:1-1} and section \ref{sec:1-2}, we introduce some background related to KMY model and Reissner-Nordstrom-Vaidya metric (RNV metric). RNV metric describes a radiating charged black hole, and is crucial in the formulation of our model. 

In section \ref{sec:2}, we give out the formulation of the thin-shell model, and work out the set of equations of motions of the sphere (\ref{eq:2-1}, \ref{eq:2-2}). We analyze this set of equations in \ref{sec:2-1}, and find out that under suitable conditions, the sphere will only move for a finite small distance in an infinite amount of time for a distant observer, implying that the black hole will not become extremal as the sphere crosses the event horizon. 

Then, we turn to the case of continuously collapsing matter in \ref{sec:3}. In \ref{sec:3-1}, we give out the basic formulation and write down the metric of the such a system (\ref{eq:3-7}), and in \ref{sec:3-2}, we work out the flux formula of the black hole (\ref{eq:3-8}, \ref{eq:3-9}). In section \ref{sec:4}, we use the equations we derive from the previous section to gain some insight of the asymptotic behavior of the black hole. 

In section \ref{sec:4-1}, we consider the simpler case in which the black hole is not near-extremal and have an extremal center followed by the  collapsing of a neutral outer layer. Based on a few assumptions (most importantly, the quasi-static approximation), the evaporation equation (\ref{eq:4-7}) can be deduced. It is further analyzed and found that this equation is only valid under the condition $x \gg \ell_{p}^{4} \slash e_{\alpha}^{3}$, where $x = a_{0} - e_{0}$. Therefore, the approximation can be applied to almost all ranges of $a_{0}$. In \ref{sec:4-2}, we turn to focus on the near-extremal case, when the criteria of the previous section does not hold, but every shell is near its extremal limit. Therefore, we can apply perturbation to the model. We show that we can deduce some properties of the asymptotic behavior by only considering the null condition $
\dot{r}_{\alpha} = -\frac{1}{2}_{\alpha}(u_{\alpha}, r_{\alpha})$. Also, by assuming the weak energy condition and some suitable initial conditions, we can get an estimation of the redshift factor $\tilde{\psi}_{\alpha} := \log (dU/du_{\alpha})$ (\ref{eq:4-19}), and we find that it will approach to $-\infty$ as $u_{\alpha}$ approaches to infinity. Also, the general Penrose diagram is given in Fig. \ref{fig:1-2} and Fig. \ref{fig:1-1}.  

In \ref{sec:4-3}, we calculate the Hawking radiation of the black hole based on the result we derived previously, and find that the Hawking radiation is always small, which justifies our assumption of chargeless field. In \ref{sec:4-4}, we give out the energy-momentum tensor of the model of the black hole. We observe that it can be decomposed into an outgoing radiation part, an infalling matter part, an electromagnetic part and a tangential part. 

\subsection{KMY Model}\label{sec:1-1}
\paragraph{ }
In the thin-shell KMY model, we consider a sphere collapsing from the past null-like infinity at the speed of light. Suppose that at any moment, the geometry outside the sphere of radius $R(u)$ can be described as the outgoing Vaidya metric:\cite{ kawai2013self,vaidya1951gravitational}
\begin{equation}
ds^{2} = -\left(1-\frac{2a(u)}{r} \right)du^{2}-2dudr +r^{2}d\Omega^2, \textrm{ where $r \geq R(u)$}
\end{equation}
which corresponds to a black hole emitting massless radiation. The energy-momentum tensor is
\begin{equation}\label{eq:a}
  T_{uu} =-\frac{\dot{a}(u)}{4\pi G r^{2}}
\end{equation}
with all the other components being equal to zero.
On the other hand, the geometry is flat inside the sphere:
\begin{equation}
ds^{2} = -dU^{2}-2dUdr +r^{2}d\Omega^2, \textrm{ where $r < R(u)$}
\end{equation}

Imposing the continuous relation between the coordinate between the inside and the outside of the sphere and the condition of the null-matter, one can obtain:
\begin{equation}\label{eq:b}
dU = -2dR = \left(1-\frac{2a(u)}{R(u)}\right)du
\end{equation}
Thus, the relationship between $u$ and $U$ and the trajectory of $R(u)$ can be found.

In order to find out the effect of Hawking radiation on the metric, one has to find the radiation flux $J$, or more precisely, the component $\langle T_{uu} \rangle$. In \cite{kawai2013self}, they calculated $J$ via Eikonal approximation 
\begin{equation}\label{eq:c}
J(u) = 4\pi r^2 \langle T_{uu} \rangle = \frac{\hbar}{8\pi}\left(\frac{\ddot{U}(u)^{2}}{\dot{U}(u)^{2}}-\frac{2\dddot{U}(u)}{3\dot{U}(u)}\right)
\equiv\ \frac{\hbar}{8\pi} \{u, U \}
\end{equation}
Thus,
\begin{equation}\label{eq:1-1}
\frac{da}{du} = -GJ = -\frac{\ell_{p}^{2}}{8\pi}\left(\frac{\ddot{R}(u)^{2}}{\dot{R}(u)^{2}}-\frac{2\dddot{R}(u)}{3\dot{R}(u)}\right)
\end{equation}
The first equality in (\ref{eq:b}) was used to replace $U(u)$ with $R(u)$ in the second equality in (\ref{eq:1-1}).

Combining the second equality of (\ref{eq:b}) and (\ref{eq:1-1}), we can solve $R(u), a(u)$ in a self-consistent way, and hence the geometry and the dynamics of the shell are found.

In \cite{kawai2013self}, they found out that $a(u)$ approaches an asymptotic value as $u$ approaches to infinity, forming an event horizon \cite{ho2015comment,kawai2013self}.

In the generic spherical symmetric case, the geometry is considered to be outgoing Vaidya between layers with their own coordinates $u_{n}$, requiring every coordinate of the layers to satisfy continuity at their own junctions. One can obtain the transformation law between the coordinate of different layers. More precisely,
\begin{equation}
-\left(1-\frac{2a_{n}(u_{n})}{R_{n}}\right)du_{n}^2-2du_{n}dR_{n} = -\left(1-\frac{2a_{n+1}(u_{n+1})}{R_{n}}\right)du_{n+1}^2-2du_{n+1}dR_{n}
\end{equation}
where $R_{n}$ (or $R_{n+1}$) is the radius of $n$th (or $n+1$th) sphere and $a_{n}$ (or $a_{n+1}$) is the Schwarzschild radius of the corresponding sphere.
In the continuous limit, the index $n$ becomes a continuous parameter $\alpha$. In \cite{kawai2013self}, the relation between $u_{\alpha_{1}}$ and $u_{\alpha_{2}}$ is calculated:
\begin{equation}\label{eq:d}
\frac{du_{\alpha_{1}}}{du_{\alpha_{2}}}= \exp\left( 2 \int_{\alpha_{2}}^{\alpha_{1}}\frac{da_{\alpha}}{R_{\alpha}(u_{\alpha})-2a_{\alpha}(u_{\alpha})}\right)
\end{equation}

The Hawking radiation at each sphere also governs the evolution of the sphere:
\begin{equation}\label{eq:f}
\frac{da_{\alpha}}{du_{\alpha}} = -\frac{\ell_{p}^{2}}{8\pi}\{u_{\alpha},U\}
\end{equation}
$U$ is the flat coordinate inside the innermost sphere. Also, by inserting (\ref{eq:d}) into (\ref{eq:f}), and using the null condition on each shell, one can obtain the evolution of the Schwarzschild radius of each shell \cite{ho2015comment,kawai2013self} :
\begin{equation}
\frac{da_{\alpha}}{du_{\alpha}} =-\frac{\ell_{p}^{2}}{384\pi a_{\alpha}^2}+O(a_{\alpha}^{-4})
\end{equation}
Therefore, the black hole will evaporate at about the time order of $a^{3}$, where $a$ is the initial Schwarzschild radius of the black hole.

\subsection{Reissner-Nordstrom-Vaidya Black Hole} \label{sec:1-2}
\paragraph{ }
In order to generalize KMY model to describe charged black holes, one needs to find an appropriate metrics to describe a charged black hole simultaneously emitting radiation. In a paper of Ibohal \cite{boulware1976hawking,ibohal2002variably,ibohal2005rotating,ibohal2010charged,wang1999generalized}, a series of metrics are proposed by letting the mass or charge parameter of Reissner-Nordstrom metric or Kerr metric to have a dependence on $u$ and $r$. The paper suggested that the dependence of those parameters contributes to the charge and mass loss of the black hole by Hawking radiation.

In particular, we will focus on one specific type of metric called Reissner-Nordstrom-Vaidya metric (RNV metric), which can be considered as a charged black hole emitting massless and chargeless radiation:\cite{ ibohal2005rotating,ibohal2010charged}

\begin{equation}\label{eq:1-3}
ds^{2} = -\left(1-\frac{2a(u)}{r}+ \frac{e^2}{r^2}\right)du^2 - 2dudr + r^2d\Omega^2
\end{equation}
where $a(u)$ and $e$ represent its mass and charge respectively.
The energy-momentum tensor of RNV black holes is:
\begin{equation}
T_{ab}=  \mu l_{a}l_{b}+2\rho l_{(a}n_{b)}+2pm_{(a}\bar{m}_{b)}
\end{equation}
The energy-momentum tensor can be decomposed into the radiation part and the electromagnetic field part:
\begin{eqnarray}
  T^{(n)}_{ab} &=& \mu l_{a}l_{b} \\
  T^{(E)}_{ab} &=& 2\rho l_{(a}n_{b)}+2pm_{(a}\bar{m}_{b)}
\end{eqnarray}
where $\mu$ can be considered as the null radiation density, and $\rho$ and $p$ are the energy density and pressure of the electromagnetic field respectively.
\begin{equation}
\mu = -\frac{\dot{a}(u)}{4\pi Gr^{2}}, \textrm{ and } p = \rho = \frac{e^2}{8\pi G r^{4}}
\end{equation}
$l_{a}$ and $n_{a}$ are real null vectors; $m_{a}$ are complex null vectors, which satisfy $n_{a}l^{a} = -1 = -m_{a}\bar{m}^{a}$
\begin{eqnarray}\label{eq:1-2}
  l_{a} &=& \delta_{a}^{0} \\
  n_{a} &=&  \frac{1}{2}\left(1-\frac{2a(u)}{r}+ \frac{e^2}{r^2}\right) \delta_{a}^{0}+ \delta_{a}^{1} \\
  m_{a} &=& -\frac{r}{\sqrt{2}}\left( \delta_{a}^{2}+ i\sin\theta \delta_{a}^{3}\right)
\end{eqnarray}

With the radiation part of the energy-momentum tensor being the same as that of Vaidya metric, we will only equate this part ($T^{(n)}_{\mu\nu}$) with the quantum field expectation value $ \langle T_{\mu\nu} \rangle$ of a massless scalar field, assuming that the electromagnetic field can be treated classically, thus it has no contribution to Hawking radiation.

One may wonder whether the radiation without charge and mass is a good approximation to this problem. In more physical cases, all charged particles have mass; if the Hawking radiation is small, the flux of particles with mass will be extremely small, thus only a few charges will evaporate from the black hole.
Therefore, it is expected that the assumption of chargeless and massless radiation is reasonable at least at the early stage of evaporation. We will find out later that Hawking radiation is low in all stages of evaporation, hence the chargeless and massless assumption is reasonable.

\section{The Collapse of a Thin-shell RNV Black Hole}\label{sec:2}
\paragraph{ }
We shall now consider a single infalling null-like sphere which is charged and collapses from the past null-like infinity. The geometry can be described as RNV metric outside the sphere and flat Minkowski inside the sphere.
\begin{equation}
ds^2 =\begin{cases}
  -dU^2-2dUdr+r^2d\Omega & \mbox{when } r < r_{s}(u) \\
  -F(u, r)du^2 -2dudr + r^{2}d\Omega^{2}  & \mbox{when } r \geq r_{s}(u)
\end{cases}
\end{equation}
$r_{s}(u)$ is the radius of the sphere, and $F(u, r) = 1-2a(u)/r+e^2/r^2$. Then, imposing the continuity condition between the inside and the outside of the sphere and the light-like condition of the sphere, we get
\begin{equation}\label{eq:2-1}
\frac{dr_{s}}{du} = -\frac{1}{2}\frac{dU}{du} = -\frac{1}{2}F(u,r_{s})
\end{equation}

If $|\dot{a}| \ll |\dot{r}_{s}|$, $a$ can be regarded as a constant and $r_{s}$ approaches $ r_{h}$ in the time order of $a$, in which $r_{h} = a +\sqrt{a^{2}-e^{2}}$ is the apparent horizon of the black hole. Using (\ref{eq:2-1}), one can get:
\begin{equation}\label{eq:2-5}
(1+2\dot{r}_{s})r_{s}^{2}-2ar_{s}+e^{2}=0 \textrm{, }\Rightarrow r_{s} = \frac{a+\sqrt{a^{2}-(1+2\dot{r}_{s})e^{2}}}{1+2\dot{r}_{s}}
\end{equation}
We shall assume that $|\dot{r}_{s}| \ll 1$, and then up to the first order,
\begin{equation}\label{eq:2-6}
r_{s} = r_{h}-2r_{h}\dot{r}_{s}-\frac{e^{2}\dot{r}_{s}}{\sqrt{a^{2}-e^{2}}} +O(\dot{r_{s}}^{2})
\end{equation}
We can see that $r_{s} > r_{h}$ if $ \dot{r}_{s}<0$. In the limit of chargeless black hole ($e \ll a$), the third term of the right hand side of (\ref{eq:2-6}) can be neglected. Otherwise, we need to include both the second term and the third term of (\ref{eq:2-6}). Also, it can be observed that the first order approximation only holds if $\frac{a^2-e^2}{e^2} \gg |\dot{r}_{s}|$.

Now we consider the flux formula for Hawking radiation. Since we are considering a spherical symmetric metric, the Eikonal approximation discussed in appendix A of \cite{kawai2013self} still applies.
\begin{equation}\label{eq:2-4}
J(u)=\frac{
\hbar}{8\pi}\left(\frac{\ddot{U}(u)^{2}}{\dot{U}(u)^{2}}-\frac{2\dddot{U}(u)}{3\dot{U}(u)}\right)
\end{equation}
but the relation of $U$ and $u$ is determined by (\ref{eq:2-1}). With (\ref{eq:a}) and the first equality of (\ref{eq:2-1}), we can deduce an equation that is similar to (\ref{eq:1-1}):
\begin{equation}\label{eq:2-2}
\dot{a}(u) = -\frac{\ell_{p}^{2}}{8\pi}\left(\frac{\ddot{r_{s}}(u)^{2}}{\dot{r_{s}}(u)^{2}}-\frac{2\dddot{r_{s}}(u)}{3\dot{r_{s}}(u)}\right)
\end{equation}
With (\ref{eq:2-1}) and (\ref{eq:2-2}) together, $a(u)$ and $r_{s}(u)$ can be solved in a consistent way.

\subsection{The Asymptotic Behavior of Thin-shell RNV Black Holes}\label{sec:2-1}
\paragraph{ }
We shall consider the evolution of the shell in two stages. In the early stage, where $|\dot{r}_{s}| \gg |\dot{a}|$, $a$ can be regarded as a constant. Thus, by setting $ r_{s} =r_{h} + x$ and $r_{s} \gg x$,
 \begin{equation}
 \dot{r}_{s} =-\frac{1}{2}F(u, r_{s}) =-\frac{1}{2}F^{\prime}(r_{h})x
 \end{equation}
 Solving $x$, we get
 \begin{equation}
 r_{s}(u) = r_{h}+C_{1}e^{-\frac{2u}{F^{\prime}(r_{h})}}
 \end{equation}
and $F^{\prime}(r_{h}) = 2(ar_{h}-e^{2})/r_{h}^{3}$. $\dot{r}_{s}$ decays exponentially. Eventually, in later stages the contribution of $\dot{a}$ becomes significant.

Assume that the black hole is not near extremal, and that $a(u)$, $r_{s}(u)$ changes rather smoothly, from (\ref{eq:2-6}) one can assume that:
 $r_{s} = r_{h}, \textrm{ }\dot{r}_{s} = \dot{r_{h}}, \textrm{ } \ddot{r_{s}}= \ddot{r_{h}} \cdots$

 To be more precise, when we say that $a(u)$, $r_{s}(u)$ change smoothly, we mean that $a(u)$, $r_{s}(u)$ satisfy $r_{s} \gg |r_{s}\dot{r_{s}}| \gg |r^{2}_{s}\ddot{r_{s}}|\cdots $ at late time. (It is the same for $a$.) This condition can be accomplished, for example, if $r_{s}$ is dominated by some negative power of $u$.

 Therefore, using the relation 
 \begin{equation}
 a = \frac{e^2+r_{h}^{2}}{2r_{h}}\textrm{ } \Rightarrow\textrm{ } \dot{a} = \frac{\dot{r}_{h}}{2}
 -\frac{e^{2}\dot{r}_{h}}{2r_{h}^2}
 \end{equation}
and inserting it into (\ref{eq:2-2}), we get
\begin{equation}\label{eq:2-3}
\dddot{r}_{h} = \frac{3}{2}\frac{\ddot{r}_{h}^{2}}{\dot{r}_{h}} + 6\pi\left(1-\frac{e^{2}}{r_{h}^{2}}\right)\dot{r}_{h}^{2}
\end{equation}

Consider that after some late time $u^{*}$, the equation (\ref{eq:2-3}) holds, and then we shall assume that the last term in the RHS of (\ref{eq:2-3}) could be neglected. Solving $\dddot{r}_{h} = 3\ddot{r}_{h}^{2}/2\dot{r}_{h}$, one can find,
\begin{equation}
  \dot{r}_{h} = -\frac{C_{2}}{(u-u_{0})^{2}} \textrm{ ,for } u> u^{*}
\end{equation}
where $u_{0} (< u^{*})$ and $C_{2}$ are integration constants. Therefore, the total distance that $r_{h}$ changes over an infinite time is finite:
\begin{equation}
  r_{h}(+\infty) = r_{h}(u^{*}) + \frac{C_{2}}{u^{*}-u_{0}}
\end{equation}
The total distance $r_{s}$ travelled after $u > u^{*}$ is $\frac{C_{2}}{u^{*}-u_{0}}$, while
\begin{equation}
\frac{2\dot{r}_{h}^{2}(u^{*})}{\ddot{r}_{h}(u^{*})} =\frac{C_{2}}{u^{*}-u_{0}}
\end{equation}
 This value is very small, since for $u \leq u^{*}$, $r_{s} \sim e^{-\alpha u} $. Therefore, $ \frac{C_{2}}{u^{*}-u_{0}} \sim e^{-\alpha u^{*}}$.

Thus, if $r_{h}$ is not very close to $a$ (not near extremal), $r_{h}$ will approach a certain value at late time limit. It will not become extremal before the sphere falls into the horizon.

Finally, we shall verify that the last term in the RHS of (\ref{eq:2-3}) can indeed be neglected. By inserting the solution $\dot{r}_{h} = C_{2}(u-u_{0})^{-2}$ into the first term and the second term of the equation, we find that
\begin{equation}
\frac{3}{2}\frac{\ddot{r}_{h}^{2}}{\dot{r}_{h}} \sim C_{2}(u-u_{0})^{-4} \textrm{, and }6\pi\left(1-\frac{e^{2}}{r_{h}^{2}}\right)\dot{r}_{h}^{2} \sim C_{2}^{2}(u-u_{0})^{-4}
\end{equation}
It is observed that $C_{2} = \frac{4\dot{r}_{h}^{3}(u^{*})}{\ddot{r}_{h}^{2}(u^{*})}$ will also be a small value since $r_{s}$ exponentially decays for $ u \leq u^{*}$.  Then, implying $ C_{2} \sim e^{-\alpha u^{*}}$, the second term which is proportional to $C_{2}^{2}$ is generally much smaller than the fist term (which is proportional to $C_{2}$).

Therefore, the second term can be neglected.

\section{The Continuously collapsing RNV black hole}\label{sec:3}
\subsection{The Continuous Limit}\label{sec:3-1}
\paragraph{ }
We now turn to the case of continuously infalling matter. We  consider many concentric infalling null shells, each shell to be labelled by an index $n$, with radius $R_{n}$. By denoting $R_{0} = r_{s}$ to be the radius of the outermost shell, the geometry between shell $n$ and shell $n-1$ can be described by Reissner-Nordstrom-Vaidya metric:
\begin{eqnarray}
  ds^{2}  &=& -F_{n}du_{n}^2-2du_{n}dr+r^2d\Omega^2 \\
   &=& -\left(1-\frac{2a_{n}(u_{n})}{r}+\frac{e_{n}^{2}}{r^{2}} \right)du_{n}^{2}-2du_{n}dr +r^{2}d\Omega^2
\end{eqnarray}
 $a_{n}(u_{n})$ generally describes the Bondi mass that is inside the $n$th shell, and $e_{n}$ is the charge inside the shell which is constant if we assume that the radiation is chargeless, and the charge is fixed on each shell.
The continuity condition between shells implies,
\begin{equation}
-F_{n}(R_{n}, u_{n})du_{n}^{2}-2du_{n}dR_{n} = -F_{n+1}(R_{n}, u_{n+1})du_{n+1}^{2}-2du_{n+1}dR_{n}
\end{equation}
 or,
 \begin{eqnarray}
    \frac{du_{n+1}}{du_{n}} &=& \frac{1}{F_{n+1}}\left[\sqrt{\left(\frac{dR_{n}}{du_{n}}\right)^{2}+F_{n+1}\left(F_{n}+2\frac{dR_{n}}{du_{n}}\right)}
    -\frac{dR_{n}}{du_{n}} \right]\\
    &=& 1-\frac{dF_{n}}{2(\dot{R_{n}}+F_{n})} +O(dF^{2}_{n})
 \end{eqnarray}
where $\frac{dR_{n}}{du_{n}} = \dot{R_{n}}$, and
\begin{equation}
dF_{n} = F_{n+1}(u_{n+1}, R_{n})-F_{n}(u_{n}, R_{n})
\end{equation}
Therefore, for $m > n$
\begin{equation}
   \frac{du_{m}}{du_{n}} =\prod_{k=n}^{m-1}\left(1-\frac{dF_{k}}{2(\dot{R_{k}}+F_{k})}\right)
\end{equation}
In the continuous limit, we change the parameter from $n$ to $\alpha$.
\begin{equation}\label{eq:3-4}
\frac{du_{\alpha_{1}}}{du_{\alpha_{2}}} = \exp\left(-\frac{1}{2}\int_{\alpha_{2}}^{\alpha_{1}}
\frac{dF_{\alpha}}{d\alpha}\frac{d\alpha}{F_{\alpha}+\dot{R_{\alpha}}}\right)
\end{equation}
We adapt the convention that $u_{0} = u$ is the retarded time of the outermost layer, and $u_{\alpha_{\textrm{max}}} =U$ is the retarded time of the innermost layer, which can also be considered as the retarded time from the past null infinity \cite{ho2016asymptotic}. For later convenience, we rewrite the index $\alpha$ in terms of r. That is:
\begin{equation}\label{eq:3-1}
R_{\alpha}(u_{\alpha}(u)) = r
\end{equation}
Using (\ref{eq:3-1}), we can redefine $\alpha$ in terms of $u$ and $r$. Thus,
\begin{eqnarray}
  F_{\alpha}(u_{\alpha}, r) &\equiv& F(u, r) = 1-\frac{2a(u,r)}{r}+\frac{e^{2}(u,r)}{r^2}\\
  \dot{R}_{\alpha} &\equiv& V(u,r)
\end{eqnarray}

The redshift factor from $r$ to $r_{s}$ is
\begin{equation}\label{eq:3-2}
\frac{du_{\alpha}}{du}  \equiv \exp \psi(u,r)
\end{equation}
\begin{eqnarray}
  \psi(u,r) &=& \frac{1}{2}\int^{r_{s}(u)}_{r}\frac{dF}{F(u, r^{\prime})+V(u, r^{\prime})} \\
   &=&\int^{r_{s}(u)}_{r}\frac{ee^{\prime}-a^{\prime}r^{\prime}}{(1+V)r^{\prime 2}-2ar^{\prime}+e^2}dr^{\prime}\label{eq:3-11} \end{eqnarray}\label{eq:3-3}
where $dF = -\frac{2da}{r^{\prime}}+\frac{2ede}{r^{\prime 2}}$ and $a^{\prime} = \frac{\partial a}{\partial r^{\prime}}$, $e^{\prime} = \frac{\partial e}{\partial r^{\prime}}$. 
If the infalling matter is null, $V(u,r)= -\frac{1}{2}F(u,r)$. Thus,
\begin{equation} \label{eq:3-13}
\psi(u, r)=2\int^{r_{s}(u)}_{r}\frac{ee^{\prime}-a^{\prime}r^{\prime}}{r^{\prime 2}-2ar^{\prime}+e^2}dr^{\prime}
\end{equation}

Then, the full metric of the system is obtained:
\begin{equation}\label{eq:3-7}
ds^2 =\begin{cases}
   -F(u,r)e^{2\psi}du^{2}-2e^{\psi}dudr+r^{2}d\Omega^{2} & \mbox{when } r < r_{s}(u) \\
  -F_{0}(u,r)du^2 -2dudr + r^{2}d\Omega^{2}  & \mbox{when } r \geq r_{s}(u)
\end{cases}
\end{equation}
where
\begin{equation}
F_{0}(u,r) = 1-\frac{2a_{0}(u)}{r}+\frac{e_{0}^2}{r^{2}}
\end{equation}
$a_{0}$ and $e_{0}$ are respectively the total Bondi mass and the total electrical charge.

\subsection{Hawking Radiation}\label{sec:3-2}
\paragraph{ }
Let $f_{\alpha}(u_{\alpha})$ be any function of $u_{\alpha}$, we can construct the ``covariant derivative'' of $f_{\alpha}$:
\begin{eqnarray}
  \frac{df_{\alpha}}{du_{\alpha}} &=& \frac{du}{du_{\alpha}}\frac{\partial f(u,r)}{\partial u}+ \frac{dr}{du_{\alpha}}\frac{\partial f(u,r)}{\partial r}\\
  &=& e^{-\psi(u,r)}\frac{\partial f(u,r)}{\partial u}+V(u,r)\frac{\partial f(u,r)}{\partial r}
\end{eqnarray}
The operation
\begin{equation}
\frac{d}{du_{\alpha}} \rightarrow D \equiv e^{-\psi}\frac{\partial}{\partial u} + V\frac{\partial}{\partial r}
\end{equation}
is called the covariant derivative with respect to the moving shells.

In order to find the evolution of each shell, note that (\ref{eq:2-2}) can be modified to be
\begin{equation}\label{eq:3-5}
\frac{da_{\alpha}}{du_{\alpha}}=-\frac{\ell_{p}^{2}}{8\pi}\{u_{\alpha},U \}
\end{equation}
$U$ is the light cone parameter from the past infinity\cite{ho2016asymptotic}, which can be identified as the retarded time of the innermost shell. By (\ref{eq:3-4}), the relation between $u_{\alpha}$ and $U$ is
\begin{equation}\label{eq:3-6}
\frac{dU}{du_{\alpha}} = e^{\tilde{\psi}(u,r)}
\end{equation}
where
\begin{equation}\label{eq:3-12}
\tilde{\psi}(u,r) = \psi(u,0) -\psi(u, r) = \int^{r}_{0}\frac{ee^{\prime}-a^{\prime}r^{\prime}}{(1+V)r^{\prime 2}-ar^{\prime}+e^2}dr^{\prime}
\end{equation}
By (\ref{eq:3-5}) and (\ref{eq:3-6}), the equation of the dynamics of $a(u,r)$ is
\begin{equation}\label{eq:3-8}
Da(u,r) = -\frac{\ell_{p}^{2}}{24\pi}\left[(D\tilde{\psi}(u,r))^{2}-2D^{2}\tilde{\psi}(u,r)\right] \textrm{, }r<r_{s}(u)
\end{equation}
In particular, if we consider the evolution of the outermost shell,
\begin{equation}\label{eq:3-9}
\dot{a}_{0}(u) =  -\frac{\ell_{p}^{2}}{24\pi}\left[\dot{\tilde{\psi}}_{0}(u,r)^{2}-2\ddot{\tilde{\psi}}_{0}(u,r)\right]\textrm{, }r\geq r_{s}(u)
\end{equation}
in which $a_{0}$ is the Bondi mass of the whole black hole, and
\begin{equation}\label{eq:3-14}
\tilde{\psi}_{0}(u) = \int^{r_{s}(u)}_{0}\frac{ee^{\prime}-a^{\prime}r^{\prime}}{(1+V)r^{\prime 2}-2ar^{\prime}+e^2}dr^{\prime}
\end{equation}

Due to the conservation of charges on each shell,
\begin{equation}\label{eq:3-10}
\frac{de_{\alpha}}{du_{\alpha}} = 0 \textrm{ } \Rightarrow De(u, r) = 0
\end{equation}

At this point, if one imposes the null matter condition of $V(u,r) = -\frac{1}{2}F(u,r)$, the geometry and the dynamics will be completely determined in a consistent way by (\ref{eq:3-11}) (\ref{eq:3-7}), (\ref{eq:3-12}), (\ref{eq:3-8}) and (\ref{eq:3-10}).

\section{The Asymptotic Analysis}\label{sec:4}
\paragraph{ }
In this section, we analyze the behavior of continuously collapsing RNV black hole under various circumstances. As an assumption, every shell obeys the null condition $\dot{r}_{\alpha} = -\frac{1}{2}F_{\alpha}(u_{\alpha},r_{\alpha})$. We will consider the case in which every collapsing shell is near its apparent horizon.  The shell being near its apparent horizon implies that $|\dot{r}_{\alpha}| \ll 1$ and the dependence of time in $a_{\alpha}(u_{\alpha})$ cannot be ignored. 

The discussion will be divided into two parts. The first one is the non-extremal case in which $a_{0}$ is not too close to $e_{0}$, in which (\ref{eq:2-6}) can hold. We also consider a region of near extremal inner layers covered by a region of collapsing neutral outer layers. Under this situation, quasi-static approximation can be applied. 

In the second part, we consider the near-extremal case, in which every shell is near its extremal limit in the initial condition. In this case, (\ref{eq:2-6}) does not hold. A more sophisticated treatment is needed under this situation. 

\subsection{The Non-extremal Behavior}\label{sec:4-1}
\paragraph{ }
Being parallel to the notation used in \cite{kawai2013self}, we define
\begin{equation} \label{eq:4-22}
\xi(u) := \frac{d}{du}\log(\frac{dU}{du}) = \dot{\tilde{\psi_{0}}} .
\end{equation}
By inserting (\ref{eq:3-14}), the definition of $\tilde{\psi}_{0}$, into (\ref{eq:4-22}) with $V = -\frac{1}{2}F$, we get

\begin{equation} \label{eq:4-1}
\xi = \frac{\dot{r_{s}}}{F}\left( \frac{-2a^{\prime}(u, r_{s})}{r_{s}} + \frac{2e^{\prime}(u, r_{s})e_{0}}{r_{s}^{2}}\right)+2\int^{r_{s}(u)}_{0}\frac{d}{du}\left( \frac{ee^{\prime}-a^{\prime}r}{r^{2}-2ar+e^{2}}\right)dr.
\end{equation}

$a^{\prime}$ denotes $\frac{\partial a}{\partial r}$, and $e^{\prime}$ is $ \frac{\partial e}{\partial r}$. The right hand side of (\ref{eq:4-1}) consists of two terms. The first term of (\ref{eq:4-1}) can be regarded as the``surface term'' which contributes to Hawking radiation emitted due to the curvature on the surface of the collapsing star, and the second term (\ref{eq:4-1}) can be considered as the ``internal term'', which contributes to Hawking radiation emitted due to the geometry inside the collapsing star. 

It is observed that the redshift from the internal of the collapsing star to the external observer is very large, since each layer is close to its apparent horizon. If some Hawking radiation is emitted from an internal shell, it is expected to experience huge redshift. This means that the energy flux from the internal shells is small compared to the outermost shell, especially when we set the initial condition of inner shells to be already near extremal. Therefore, the internal term in (\ref{eq:4-1}) can be neglected, by assuming the surface term to be much larger. 

To be more percise, we adapt the quasi-static approximation. The quasi-static approximation means that, due to the huge redshift, the inner shells ``freeze'' when compared to the outermost shell. $a(u, r)$ can be seen as solely dependent on $r$ for $r < r_{s}$. Therefore,

\begin{equation}\label{eq:4-10}
a(u,r) =\begin{cases}
   a(r) & \mbox{when } r < r_{s}(u) \\
  a_{0}(u)  & \mbox{when } r \geq r_{s}(u)
\end{cases}.
\end{equation}

In order to calculate $\xi$, we have to impose some boundary condition for $a(u, r)$ and $e(u,r)$ on the junction $r = r_{s}(u)$. Namely, we have to find $a^{\prime}(u, r_{s})$ and $e^{\prime}(u,r_{s})$.

To find $a^{\prime}(u, r_{s})$, the equation of Hawking radiation (\ref{eq:3-8}), (\ref{eq:3-9}) should be consistent. Therefore, the continuity condition of $Da(u,r)$ on $r=r_{s}$ can be imposed, which can be used to find $ a^{\prime}(u, r_{s})$, 

\begin{equation}\label{eq:4-2}
\dot a_{0}(u) = \lim_{r \to r_{s}-} Da(u,r) = \dot{r_{s}} \left.\frac{\partial a}{\partial r}\right|_{r=r_{s}}.
\end{equation}

To find $e^{\prime}(u, r_{s})$, we shall assume that $e^{\prime}(u, r_{s}) = 0$. Since the charge is fixed on each shell, this can be achieved by manipulating the initial distribution of charge on each shell. For example, it can be done by letting a region of the outermost shells contain no charge. 

(\ref{eq:4-2}) can give the form of $a(r)$ in the region of $r$ where the quasi-static approximation can apply. More precisely, the form of $a(r)$ can be solved by the equation
\begin{equation} \label{eq:4-3}
a^{\prime}(r_{s}(u)) = \frac{\dot{a}_{0}}{\dot{r}_{s}} = \frac{\sqrt{a_{0}^{2}-e_{0}^{2}}}{r_{h}}.
\end{equation}

We have adopted the assumption that the shells are near their apparent horizon in the second equality of (\ref{eq:4-3}). 
Therefore, for the quasi-static approximation to hold, a specific initial condition of $a(r, u_{0})$, which is the solution of (\ref{eq:4-3}), have to be set. $u_{0}$ is some initial time such that each shell is sufficiently close to its apparent horizon for $u \geq u_{0}$. 

Under this circumstance, the configuration of $a(u, r)$ and $e(u, r)$ corresponds to neutral outer layer in the region of $r_{0} \leq r \leq r_{s}(u)$. $r_{0}$ is the minimum of $r_{s}(u)$ such that the quasi-static approximation holds.  Since each shell is close to its horizon, $e(u, r)= \sqrt{2a(u, r)r-r^{2}}$. It implies that $e(u, r)$ is also independent of $u$  for $r < r_{s}$. We have set the boundary condition to $e^{\prime}(u, r_{s}) = 0$.  Therefore, we conclude that $e(u, r) = e_{0}$ for $r \geq r_{0}$. Since we are assuming that the collapsing star has a near extremal inner shells, we have $r_{0} \approx e_{0}$.

We precede to evaluate $\xi(u,r)$. Since the second term of (\ref{eq:4-1}) and $e^{\prime}(u, r_{s})$ vanish under our assumptions,

\begin{equation}
\xi(u, r)  = -\frac{1}{F(u,r_{s})}\frac{2\dot a_{0}}{r_{s}}.
\end{equation}

Since the non-extremal condition (\ref{eq:2-6}) holds, we can use the assumptions  $r_{s} = r_{h}$ and $\dot{r}_{s} = \dot{r}_{h}$. The latter assumption holds if $r_{s}$ changes rather smoothly as discussed in section \ref{sec:2-1}. With the null condition $F(u, r_{s}) = -\frac{1}{2}\dot{r_{s}}$, $F(u, r_{s})$ can be rewritten as 

\begin{equation}\label{eq:4-20}
F(u, r_{s}) = -\frac{2r_{h}\dot{a}_{0}}{\sqrt{a_{0}^{2}-e_{0}^{2}}}.
\end{equation}
By inserting (\ref{eq:4-20}) into $\xi$, we get
\begin{equation}\label{eq:4-4}
\xi=-\frac{1}{F(u,r_{s})}\frac{2\dot a_{0}}{r_{s}}=\frac{\sqrt{a_{0}^{2}-e_{0}^{2}}}{r_{h}^{2}}.
\end{equation}
In order to find the evolution of $a_{0}$, we need to recall the previous result (\ref{eq:3-9})
\begin{equation}\label{eq:4-5}
\dot{a}_{0} = \frac{\ell_{p}^{2}}{24\pi}(2\dot{\xi}-\xi^{2}).
\end{equation}
Inserting the relation of $\xi$ (\ref{eq:4-4}) into the above, the evaporation equation of $a_{0}$ can be derived:
\begin{equation}\label{eq:4-6}
\dot{a}_{0} =-\frac{\ell_{p}^{2}}{24\pi}\left(
1+\frac{\ell^{2}_{p}}{12\pi r_{h}^{2}}\left(2-\frac{a_{0}}{\sqrt{a_{0}^{2}-e_{0}^{2}}}\right)\right)^{-1}\frac{a_{0}^{2}-e_{0}^{2}}{r_{h}^{4}}
\end{equation}
For $a_{0}$ which is not too close to $e_{0}$, (\ref{eq:4-6}) can be simplified to
\begin{equation}\label{eq:4-7}
\dot{a}_{0} =-\frac{\ell_{p}^{2}}{24\pi}\frac{a_{0}^{2}-e_{0}^{2}}{r_{h}^{4}}.
\end{equation}

It can be observed that, in the limit $a_{0} \gg e_{0}$, the equation becomes $\dot{a}_{0} = -\ell_{p}^{2}/(24\pi a_{0}^{2})$, which is consistent with the result derived by Kawai et al.\cite{kawai2013self}

On the other hand, if we let $a_{0} \rightarrow e_{0}$ in (\ref{eq:4-6}), $\dot{a_{0}}$ becomes positive. It implies an unnatural negative energy flux, but it should be noted that we have made the non-extremal assumption which may not hold in the late time limit when deriving the result. Therefore, if we prove that $\dot{a}_{0}$ is always positive under the range of $a_{0}$ where (\ref{eq:2-6}) holds, the problem can be resolved.

Actually, we can show that the evaporation equation (\ref{eq:4-7}) and the non-extremal condition (\ref{eq:2-6}) are consistent under the same condition $\frac{\ell_{p}^{4}}{e_{0}^{3}} \ll x$ where $a_{0} = e_{0}+ x$.

To see this, we begin from the fact that (\ref{eq:2-6}) holds if and only if  $\frac{a_{0}^2-e_{0}^2}{e_{0}^2} \gg |\dot{r}_{s}|$. Under the condition in which (\ref{eq:4-7}) holds, $\dot{r}_{s}$ can be rewritten as

\begin{equation}\label{eq:4-8}
\dot{r}_{s} \approx \dot{r}_{h} = \frac{r_{h}\dot{a_{0}}}{\sqrt{a_{0}^{2}-e_{0}^{2}}}= -\frac{\ell_{p}^{2}}{24\pi}\frac{\sqrt{a_{0}^{2}-e_{0}^2}}{r_{h}^{3}}.
\end{equation}
Letting $a_{0} = e_{0} + x$ and $x \ll e_{0}$, the condition turns out to be
\begin{equation}\label{eq:4-9}
 |\dot{r}_{s}| \ll \frac{a_{0}^2-e_{0}^2}{e_{0}^2} \textrm{ } \Rightarrow \sqrt{\frac{\ell^{4}_{p}x}{e_{0}^{5}}} \ll \frac{x}{e_{0}} \textrm{, or } \frac{\ell_{p}^{4}}{e_{0}^{3}} \ll x.
 \end{equation}
 We should now check the consistency with (\ref{eq:4-7})
\begin{equation}
\frac{a_{0}\ell_{p}^{2}}{\sqrt{a_{0}^{2}-e_{0}^{2}}r_{h0}^{2}} \approx \sqrt{\frac{\ell_{p}^{4}}{e_{0}^{3}x}} \ll 1.
\end{equation}
Indeed, the approximation from (\ref{eq:4-6}) to (\ref{eq:4-7}) is consistent with the non-extremal assumption. We see that the valid range for (\ref{eq:4-7}) is $\frac{\ell_{p}^{4}}{e_{0}^{3}} \ll x$, which is sub-Planck scale for $e_{0} > 1$. 

\subsection{Comments on the Near-Extremal Behavior}
\label{sec:4-2}
\paragraph{ }
In the previous section, we demonstrated that, for non extremal RNV black holes, the evaporation process is governed by (\ref{eq:4-7}) under suitable assumptions: the non-extremal conditions (\ref{eq:2-6}) and the quasi-static approximation. We checked that (\ref{eq:4-7}) is valid when $\ell_{p}^{4} \slash e_{0}^{3} \ll x$, where $a_{0} = e_{0} + x$. In the late time limit, both of the conditions may not hold. Also, we found that the quasi-static assumption leads to a particular configuration of $a(u, r)$ and $e(u, r)$.

We consider a more general situation in this section. Both the quasi-static and  (\ref{eq:2-6}) are not assumed. To avoid complexity in later calculations, we will adopt a general parameterization $\alpha$ on different layers. $u_{0} = u$ is the retarded time for the outermost layer, and $u_{\alpha_{\textrm{max}}} = U$ is the retarded time for the innermost layer. 
Also, we will consider the near-extremal limit, in which each shell is very close to its extremal limit. Therefore, $r_{\alpha} \approx e_{\alpha}$ and $a_{\alpha} \approx e_{\alpha}$, and perturbation on $a_{\alpha}$ and $e_{\alpha}$ can be evoked. 

In order to analyze the asymptotic behavior of the near-extremal black hole, we need to replace (\ref{eq:2-6}) with (\ref{eq:2-5}), and the second term of (\ref{eq:4-1}) cannot be neglected. Generally speaking, the equation is highly non-linear. 
This section provides an analysis of the behavior of $a_{\alpha}$, $r_{\alpha}$, or the redshift factor $\tilde{\psi}_{\alpha}$ under different assumptions, and we give an illustration of how the general Penrose diagram of the RNV black hole looks like.

Before invoking the equation for Hawking radiation (\ref{eq:3-8}), there are many things we can say about by solely considering the null condition. The null condition for an arbitrary layer $\alpha$ is
\begin{equation}\label{eq:4-12}
\dot{r_{\alpha}} = -\frac{1}{2}F_{\alpha}(u_{\alpha},r_{\alpha}).
\end{equation}
 Since $r_{\alpha} \approx e_{\alpha}$ and $a_{\alpha} \approx e_{\alpha}$, we can make perturbation on $a_{\alpha} = e_{\alpha}+x_{\alpha}$ and $r_{\alpha} = e_{\alpha}+y_{\alpha}$. To obtain consistency, the perturbation of (\ref{eq:4-12}) needs to be taken to the second order, which is
\begin{equation}\label{eq:4-13}
\dot{y_{\alpha}} = \left(\frac{x_{\alpha}}{e_{\alpha}}\right)-\frac{1}{2}\left(\frac{y_{\alpha}}{e_{\alpha}}\right)^{2}- \left(\frac{x_{\alpha}}{e_{\alpha}}\right) \left(\frac{y_{\alpha}}{e_{\alpha}}\right) + O(x_{\alpha}y^{2}_{\alpha}).
\end{equation}
Note that we will assume there exists some late time $u_{\alpha}^{*}$ such that the perturbation (\ref{eq:4-13}) will holds for all $u_{\alpha} > u_{\alpha}^{*}$. 

Rearranging (\ref{eq:4-13}), we get
\begin{equation}\label{eq:4-14}
\frac{x_{\alpha}}{e_{\alpha}} = \frac{\dot{y_{\alpha}}+\frac{1}{2}\left(\frac{y_{
\alpha}}{e_{\alpha}}\right)^{2}}{1-y_{\alpha} \slash e_{\alpha}} \approx \dot{y_{\alpha}}+\frac{1}{2}\left(\frac{y_{
\alpha}}{e_{\alpha}}\right)^{2}.
\end{equation}
One important constraint is that we need to have $x_{\alpha} > 0$, or in other words, $\dot{y_{\alpha}}+\frac{1}{2}\left(\frac{y_{\alpha}}{e_{\alpha}}\right)^{2} > 0$. To further analyze this, we can rewrite this inequality to be
\begin{equation}
\frac{d}{du_{\alpha}}\left(-\frac{2e_{
\alpha}^{2}}{y_{\alpha}} + u_{\alpha}\right) > 0 \textrm{ } \Rightarrow -\frac{2e_{
\alpha}^{2}}{y_{\alpha}} + u_{\alpha} = f(u_{\alpha}).
\end{equation}
$f(u_{\alpha})$ is a monotonic increasing function, which is $f(u_{\alpha}) = \int \frac{2e_{\alpha}x_{\alpha}}{y_{\alpha}^{2}}du_{\alpha}$. 
Therefore, we find that
\begin{equation}\label{eq:4-15}
y_{\alpha} = \frac{2e_{\alpha}^{2}}{u_{\alpha} - f(u_{\alpha})} > \frac{2e_{\alpha}^{2}}{u_{\alpha} - f(u_{\alpha}^{*})}.
\end{equation}
If we match up the initial condition, we will get $ f(u_{\alpha}^{*}) = -u_{\alpha}^{*}+2e_{\alpha}^{2} \slash y_{\alpha}(u_{\alpha}^{*})$.
Another constraint is $\dot{y}_{\alpha} < 0$.  Since the shell is outside the apparent horizon $r_{h\alpha} := a_{\alpha}+\sqrt{a_{\alpha}^{2}-e_{\alpha}^{2}}$, from the null condition, we can see that $\dot{y}_{\alpha}$ will always remain negative ($F_{\alpha} > 0$). From (\ref{eq:4-15}), this implies 
\begin{equation}\label{eq;4-21}
\frac{2e_{\alpha}x_{\alpha}}{y_{\alpha}^{2}} = \dot{f}(u_{\alpha}) < 1.
\end{equation}

Now, we shall consider the asymptotic behavior $x_{\alpha}/y_{\alpha}^{2}$ as $u_{\alpha} \rightarrow \infty$. Suppose that $x_{\alpha}/y_{\alpha}^{2}$ approaches some limit value denoted as $\eta_{\alpha}$, and we know from (\ref{eq;4-21}) that $\eta_{\alpha} \leq 1/2e_{\alpha}$.

Then, we consider three different cases of the value of $\eta_{\alpha}$:

\begin{enumerate}[label={(\arabic*)}, noitemsep]
\item $\eta_{\alpha} = 0$ \label{case1}
\item  $0< \eta_{\alpha} < 1/2e_{\alpha}$\label{case2}
\item $\eta_{\alpha} = 1/2e_{\alpha}$\label{case3}
\end{enumerate}

\paragraph{Case (1)}
In case \ref{case1}, it can be observed that $\eta_{\alpha} = 0$ implies $\dot{f}(u_{\alpha}) \rightarrow 0$ as $u_{\alpha} \rightarrow \infty$. Therefore, for any $\epsilon > 0$ and large enough $u_{\alpha}$, the inequality $f(u_{\alpha}) < \epsilon u_{\alpha}$ can be achieved. Thus, we can expand (\ref{eq:4-15}) with respect to $f(u_{\alpha})/ u_{\alpha}$ and find that
\begin{equation}\label{eq:4-23}
y_{\alpha} = \frac{2e_{\alpha}^{2}}{u_{\alpha}}+o(u_{
\alpha}^{-1}).
\end{equation}
Also, since $x_{\alpha}/y_{\alpha}^{2} \rightarrow 0$, we can conclude that $x_{\alpha} = o(u_{\alpha}^{-2})$. 
\paragraph{Case (2)}
In case \ref{case2}, similarly, we observe that $\dot{f}(u_{\alpha}) \rightarrow 2\eta_{\alpha} e_{\alpha} < 1$ as $u_{\alpha} \rightarrow \infty$. Therefore, $f(u_{\alpha}) = 2\eta_{\alpha} e_{\alpha} u_{\alpha} + o (u_{\alpha})$, and this implies 
\begin{equation}\label{eq:4-24}
y_{\alpha} = \frac{A_{\alpha}}{u_{\alpha}}+ o(u_{\alpha}^{-1}),
\end{equation}
 where $A_{\alpha}$ is some constant which can be written explicitly as $A_{\alpha} = \frac{2e_{\alpha}^{2}}{1- 2\eta_{\alpha} e_{\alpha}}$. On the other hand, since $x_{\alpha} = \eta_{\alpha} y_{\alpha}^{2}$ for $u_{\alpha} \rightarrow \infty$,
 
\begin{equation}\label{eq:4-25}
 x_{\alpha} = \frac{B_{\alpha}}{u_{\alpha}^{2}}+ o(u_{\alpha}^{-2}),
\end{equation}
where $B_{\alpha} = \frac{4\eta_{\alpha} e_{\alpha}^{4}}{(1-2\eta_{\alpha} e_{\alpha})^{2}}$.

Also, we can observe that $A_{\alpha}$ and $B_{\alpha}$ satisfy  
\begin{equation}
B_{\alpha} = \frac{A_{\alpha}^{2}}{2e_{\alpha}}-A_{\alpha}e_{\alpha} 
\end{equation}

\paragraph{Case (3)}
In case \ref{case3}, we can only conclude that $y_{\alpha} = \sqrt{2e_{\alpha}x_{\alpha}}$ at the late time limit. By observing that the apparent horizon is just $r_{h\alpha} = a_{\alpha}+\sqrt{a_{\alpha}^{2}-e_{\alpha}^{2}} \approx e_{\alpha}+\sqrt{2e_{\alpha}x_{\alpha}}$, this is the case in which the distance between each shell and its horizon is much smaller than the distance of each shell and its extremal limit $e_{\alpha}$. Therefore, $r_{s} = r_{h}$ still holds under this situation. In this case, the evolution of $y_{\alpha}$ is fully determined by the evolution of $x_{\alpha}$, which itself is determined by the evaporation process.

It should be noted that, in case \ref{case1} and case \ref{case2}, $y_{\alpha}$ and $x_{\alpha}$ will approach to zero as $u_{\alpha} \rightarrow \infty$, but in case \ref{case3} it is not obvious whether $x_{\alpha}$ will approach to zero as $u_{\alpha} \rightarrow \infty$. 

In order to obtain more information, we will focus on the equation of evaporation (\ref{eq:3-8}). Written in the general parameterization $\alpha$, it is
\begin{equation}\label{eq:4-16}
\dot{a}_{\alpha} = \frac{\ell_{p}^{2}}{24\pi}(2\dot{\xi}_{\alpha}-\xi_{\alpha}^{2}),
\end{equation}
and $\xi_{\alpha}$ is defined as 
\begin{equation}\label{eq:4-18}
\xi_{\alpha} = \frac{d}{du_{\alpha}}\log\left(\frac{dU}{du_{\alpha}}\right)  
\end{equation}

We shall assume that the energy momentum tensor holds under the weak energy condition. This demands the total energy flux of the black hole to be positive, which corresponds to $\dot{a}_{\alpha} < 0$. From the same technique used in deriving (\ref{eq:4-15}), this criteria can be reformulate as 
\begin{equation}
2\dot{\xi}_{\alpha}-\xi_{\alpha}^{2} < 0 \Rightarrow \frac{d}{du_{\alpha}}\left( \frac{2}{\xi_{\alpha}} + u_{\alpha} \right) > 0.
\end{equation}
Therefore, if we define $g(u_{\alpha}) = -\int \frac{ 24\pi\dot{a}_{\alpha}}{\ell_{p}^{2}\xi_{\alpha}^{2}} du_{\alpha}$, which is monotonic increasing, we get
\begin{equation}
\frac{2}{\xi_{\alpha}} + u_{\alpha} = g(u_{\alpha}), \textrm{ }\Rightarrow \xi_{\alpha} = -\frac{2}{u_{\alpha}- g(u_{\alpha})} < -\frac{2}{u_{\alpha}-g(u_{\alpha}^{*})}.
\end{equation}
Matching up the initial condition, we get $g(u_{\alpha}^{*}) = u_{\alpha}^{*}+\frac{2}{\xi_{\alpha}(u_{\alpha}^{*})}$. 

In general, we can assume that $\xi_{\alpha}(u_{\alpha}^{*}) < 0$ \footnote{In general, it is natural to assume that $\xi_{\alpha} < 0$ in the late time limit. This can be seen as the following: from the definition, $\xi_{\alpha} = \frac{d^{2}U}{du^{2}_{\alpha}}/\frac{dU}{du_{\alpha}}$. $\xi_{\alpha} < 0$ implies $\frac{d^{2}U}{du_{\alpha}^{2}} < 0$. This is reasonable because one would expect the redshift between the innermost layer and $\alpha$ to increase with time due to the approaching of each layer to its apparent horizon. Also, since there are little Hawking radiation emitted in the near-extremal limit, there is no mass loss of the black hole to reduce its redshift factor. This implies that $\frac{dU}{du_{\alpha}}$ should decrease with time.}, and there is no singularity in $\xi_{\alpha}$. Therefore, an estimation of the redshift factor $\tilde{\psi}_{\alpha} := \frac{dU}{du_{\alpha}}$ can be obtained:
\begin{equation}\label{eq:4-19}
\tilde{\psi}_{\alpha}(u_{\alpha})-\tilde{\psi}_{\alpha}(u_{\alpha}^{*}) < -2\log\left(\frac{u_{\alpha}-g(u_{\alpha})}{u_{\alpha}^{*}-g(u_{\alpha}^{*})}\right)
\end{equation}
Therefore, $\tilde{\psi}_{\alpha}$ will diverge to $-\infty$ as $u_{\alpha} \rightarrow \infty$. 

In case \ref{case1} and case \ref{case2}, we have $x_{\alpha} = O(u_{\alpha}^{-2})$. Also, since that 
\begin{equation}
\frac{24\pi\dot{a}_{\alpha}}{\ell_{p}^{2}\xi_{\alpha}^{2}} =\frac{24\pi\dot{x}_{\alpha}}{\ell_{p}^{2}\xi_{\alpha}^{2}}> \frac{6\pi\dot{x}_{\alpha} }{\ell_{p}^{2}}(u_{\alpha}-g(u_{\alpha}^{*}))^{2},
\end{equation}  
we can get an estimation of the growth of $g(u_{\alpha})$:
\begin{eqnarray}
g(u_{\alpha})-g(u_{\alpha}^{*}) &<& -\frac{6\pi}{\ell_{p}^{2}}\int_{u_{\alpha}^{*}}^{u_{\alpha}}\dot{x}_{\alpha}(u^{\prime}-g(u_{\alpha}^{*}))^{2}du^{\prime} \label{eq:4-17-1}\\
&=& \left.-\frac{6\pi x_{\alpha}}{\ell_{p}^{2}}(u^{\prime}-g(u_{\alpha}^{*}))^{2}
\right|_{u_{\alpha}}^{u_{\alpha}^{*}} + \frac{12\pi}{\ell_{p}^{2}}\int_{u_{\alpha}^{*}}^{u_{\alpha}}x_{\alpha}(u^{\prime}-g(u_{\alpha}^{*}))du^{\prime}\label{eq:4-17-2}
\end{eqnarray}

The first term of (\ref{eq:4-17-2}) is $O(-1)$, and the second term is $O(\log(u_{\alpha}))$. Therefore, under the circumstances of case \ref{case1} and case \ref{case2}, $g(u_{\alpha}) = O(\log(u_{\alpha}))$. Thus, $\xi_{\alpha} = -2 / u_{\alpha} +O(\log(u_{\alpha})/u_{\alpha}^{2})$. Using (\ref{eq:4-18}), the general relationship between the retarded time of two arbitrary layers $u_{\alpha_{1}}$ and $u_{\alpha_{2}}$ can be obtained:
\begin{equation}
\frac{du_{\alpha_{1}}}{du_{\alpha_{2}}} = \frac{du_{\alpha_{1}}}{dU}\frac{dU}{du_{\alpha_{2}}}=   e^{\int \xi_{\alpha_{2}} du_{\alpha_{2}}-\int \xi_{\alpha_{1}} du_{\alpha_{1}}} = \frac{c_{\alpha_{1}}u_{\alpha_{1}}^{2}}{c_{\alpha_{2}}u_{\alpha_{2}}^{2}},
\end{equation}
in which $c_{\alpha_{1}}$ and $c_{\alpha_{2}}$ are integral constants. Integrating this result, the relation between $u_{\alpha_{1}}$ and $u_{\alpha_{2}}$ is
\begin{equation}
-\frac{1}{c_{\alpha_{1}}u_{\alpha_{1}}}+c_{\alpha_{1}}^{\prime} = -\frac{1}{c_{\alpha_{2}}u_{\alpha_{2}}}+c_{\alpha_{2}}^{\prime},
\end{equation}
in which $c_{\alpha_{1}}^{\prime}$ and $c_{\alpha_{2}}^{\prime}$ are also integral constants. If $\alpha_{2} > \alpha_{1}$ ($\alpha_{2}$ is below $\alpha_{1}$), by letting $u_{\alpha_{1}}\rightarrow \infty$, $u_{\alpha_{2}}$ approaches to its limiting value $u_{\alpha_{2}}(\infty)=\frac{1}{c_{\alpha_{2}}(c_{\alpha_{2}}^{\prime}-c_{\alpha_{1}}^{\prime})}$. For consistency, it is required that $u_{\alpha_{2}}(\infty) > u_{\alpha_2}^{*}$.

The general Penrose diagram of the RNV black hole, not limited to case \ref{case1} and case \ref{case2}, can be depicted as the following: 

\begin{figure}[!tbp]

  \centering 
  \begin{minipage}[t]{0.4\textwidth} 
    \includegraphics[width=\textwidth]{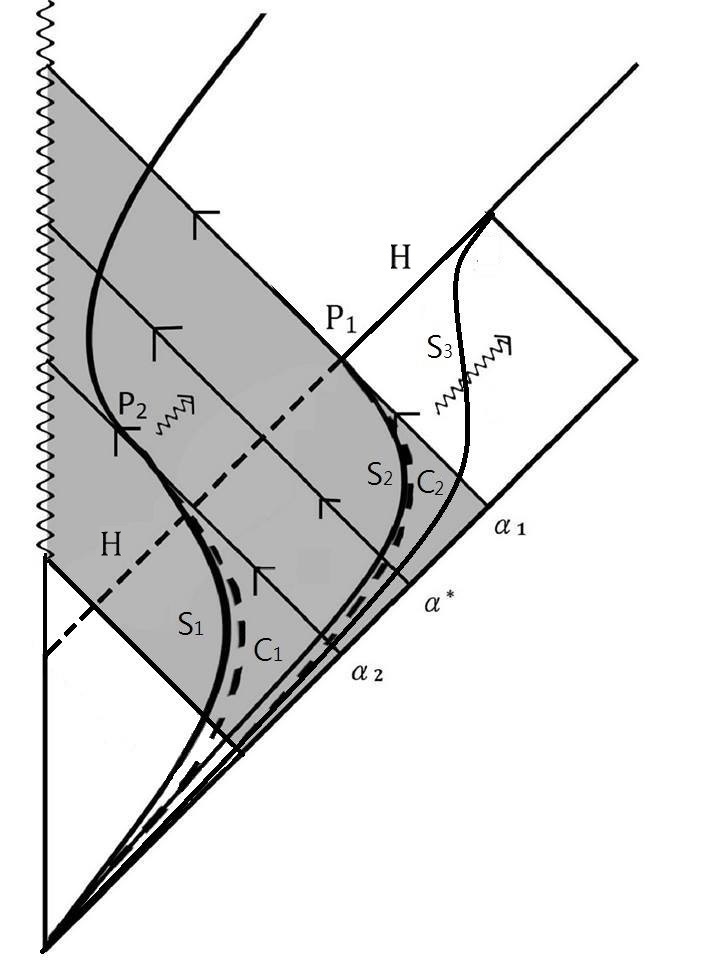}
    \caption{The Penrose diagram that exhibits the critical layer $\alpha^{*}$ is depicted as the following. The crucial difference of this figure with figure 2 is the existence of $\alpha^{*}$, such that any layer $\alpha_{2}$ which is below $\alpha^{*}$ will only become extremal inside the horizon $H$ (at $P_{2}$). Therefore, it will still emit some radiation into an asymptotic region other than the region of the observer. The solid curves are constant $r$ curves and the dashed curves represent constant $a$ curves. More precisely,   $S_{1}, S_{2}$ and $S_{3}$ respectively represent $r = e_{\alpha_{1}}$, $r = e_{\alpha_{2}}=e_{0}$ and $r = \textrm{const.} > e_{0}$. On the other hand, $C_{1}$ and $C_{2}$ respectively represent $a = e_{\alpha_{1}}$ and $ a = e_{\alpha_{2}} = e_{0}$.}
    \label{fig:1-2}  \end{minipage}
  \hfill
  \begin{minipage}[t]{0.4\textwidth}
    \includegraphics[width=\textwidth]{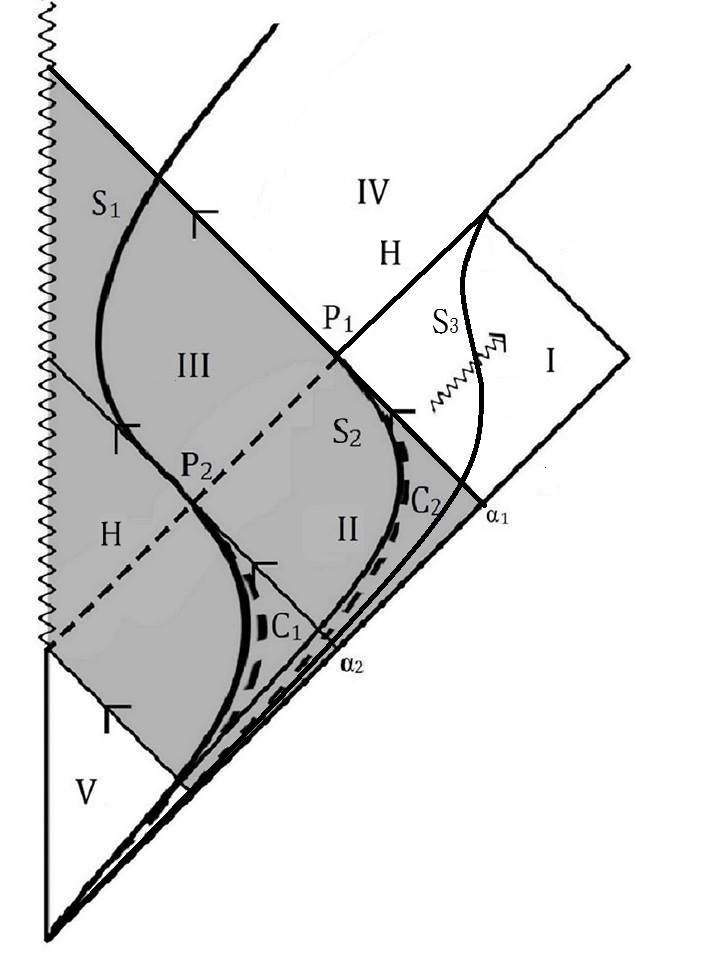}
     \caption{The Penrose diagram that does not exhibit the critical layer $\alpha^{*}$ is depicted as the following. The 
     definitions of the solid curves and dash curves are the same as in figure 1.
    This figure can be divided into five regions:
    Region I is outside the outermost layer of the black hole, containing the asymptotic flat region of our universe. Region II is the region between infalling shells, and the metric here is described as the $r<r_{s}$ part of (\ref{eq:3-7}). Region III is the part where the shell falls into the horizon $H$. All shells have become extremal; therefore, no Hawking radiation can be emitted here. Region IV is another asymptotically flat region. The metric here is the same as the ordinary RN black holes. Region V is a place where the spacetime is not influenced by the collapsing matter, which can be represented by a flat metric. }
    
    \label{fig:1-1}
  \end{minipage}
\end{figure}

Considering the outermost layer $\alpha = 0$, from the previous derived result, we know that $\tilde{\psi}_{0} = \log(\frac{dU}{du})$ will approach to $-\infty$ in the late time limit. This implies that there exists some layer $\alpha^{\prime}$ such that the redshift factor $\psi_{\alpha} : = \log(du_{\alpha}/du)$ approaches to negative infinity at late time limit for all $\alpha < \alpha^{\prime}$, and approaches to a finite value for $\alpha > \alpha^{\prime}$.

It is important to note that $\psi_{\alpha} \rightarrow -\infty$ does not necessary imply $u_{\alpha} \rightarrow \textrm{const.} < \infty$ as $u \rightarrow \infty$. For example, if $\psi_{\alpha} = -\log(u_{\alpha})$, $u_{\alpha} = \sqrt{2u+C_{\alpha}}$, which approaches infinity as $u \rightarrow \infty$.

Therefore, there are two possible cases. The first case is the existence of a critical shell $\alpha^{*} \geq \alpha^{\prime}$, such that any layer above $ \alpha^{*}$ will become extremal as the time of the observer approaches infinity; otherwise it won't. (Figure \ref{fig:1-2})

For the second case, such a critical layer does not exist. This implies that an observer with a distance to the black hole will find every layer of the shell becomes extremal as the time approaches to infinity. (Figure \ref{fig:1-1})

We can see this more clearly with the help of the Penrose diagrams (Fig.\ref{fig:1-2} and Fig.\ref{fig:1-1}). In the figures, $\alpha_{1}$ and $\alpha_{2}$ are the trajectories of two infalling null shells (thus they are also constant $e$ curves), and $S_{1}$, $S_{2}$ and $S_{3}$ are respectively the constant $r$ curves with $r = e_{\alpha_{1}}$, $r=e_{\alpha_{2}} = e_{0}$ and $r = \textrm{const.} > e_{0}$. The dash curves $C_{1}$ and $C_{2}$ are respectively the constant $a$ curves for $a = e_{\alpha_{1}}$ and $a = e_{\alpha_{2}}$. In figure \ref{fig:1-1}, we can see that $S_{1}$, $S_{2}$, $C_{1}$ and $C_{2}$ cross the null shells $\alpha_{1}, \alpha_{2}$ as they approach the horizon $H$ (at $P_{1}, P_{2}$), while in figure \ref{fig:1-2}, $S_{2}$ and $C_{2}$ cross the null shell inside the horizon $H$, meaning that an outside observer cannot see this event.

Finally, it should be noted that the analysis of this section is based on the assumption that $r_{\alpha}$ and $a_{\alpha}$ are already near-extremal for all layers initially. It is possible that in a general collapse of a charged black hole, some or all layers will not approach their extremal limit in the late time limit. For example, in section \ref{sec:2} we know that the collapsing of a single charged shell will not become extremal in the late time limit. On the other hand, the thin shell model can be simulated by taking a limit of $a(u, r)$ and $e(u, r)$ into delta functions. It is worth investigating whether this behavior (being non-extremal in late time limit) is a consequence of the delta function in $a$ and $e$, or a more general behavior.

\subsection{The Hawking Temperature}\label{sec:4-3}
\paragraph{ }
In this section, we study the Hawking temperature of the collapsing RNV black hole, subjected to the assumptions of \ref{sec:4-1} in more detail. Since the evaporation equation (\ref{eq:4-7}) is derived under the condition  $\frac{\ell_{p}^{4}}{e_{0}^{3}} \ll x$, the Hawking temperature is determined for all $a_{0}$ subjected under this condition.

We aim at finding the character of the Hawking temperature. We find that, being opposed to the chargeless case, the Hawking temperature will have a maximum at some point, and it will approach to 0 as $a_{0} \rightarrow e_{0}$ and $a_{0}\rightarrow \infty$.

According to section \ref{sec:4-1}, the energy flux density is, 
\begin{equation}
J =  \frac{1}{4\pi r_{h}^{2}}\int_{r=r_{h}} d\Omega T_{uu} = -\frac{\dot{a}_{0}}{4\pi r_{h}^{2}} = \frac{1}{96\pi^{2}}\frac{a_{0}^{2}-e_{0}^{2}}{r_{h}^{6}} 
\end{equation}
and the Hawking temperature is $T_{H}$

\begin{equation}\label{eq:106}
T_{H} = \left(\frac{J}{\sigma}\right)^{1/4} .
\end{equation}
In the chargeless limit, $e_{0} \ll a_{0}$, $T_{H} \propto 1/a_{0} $, which coincides with the calculation of Kawai et al.\cite{kawai2013self}.

\begin{figure}
  \centering
      \includegraphics[width=13cm]{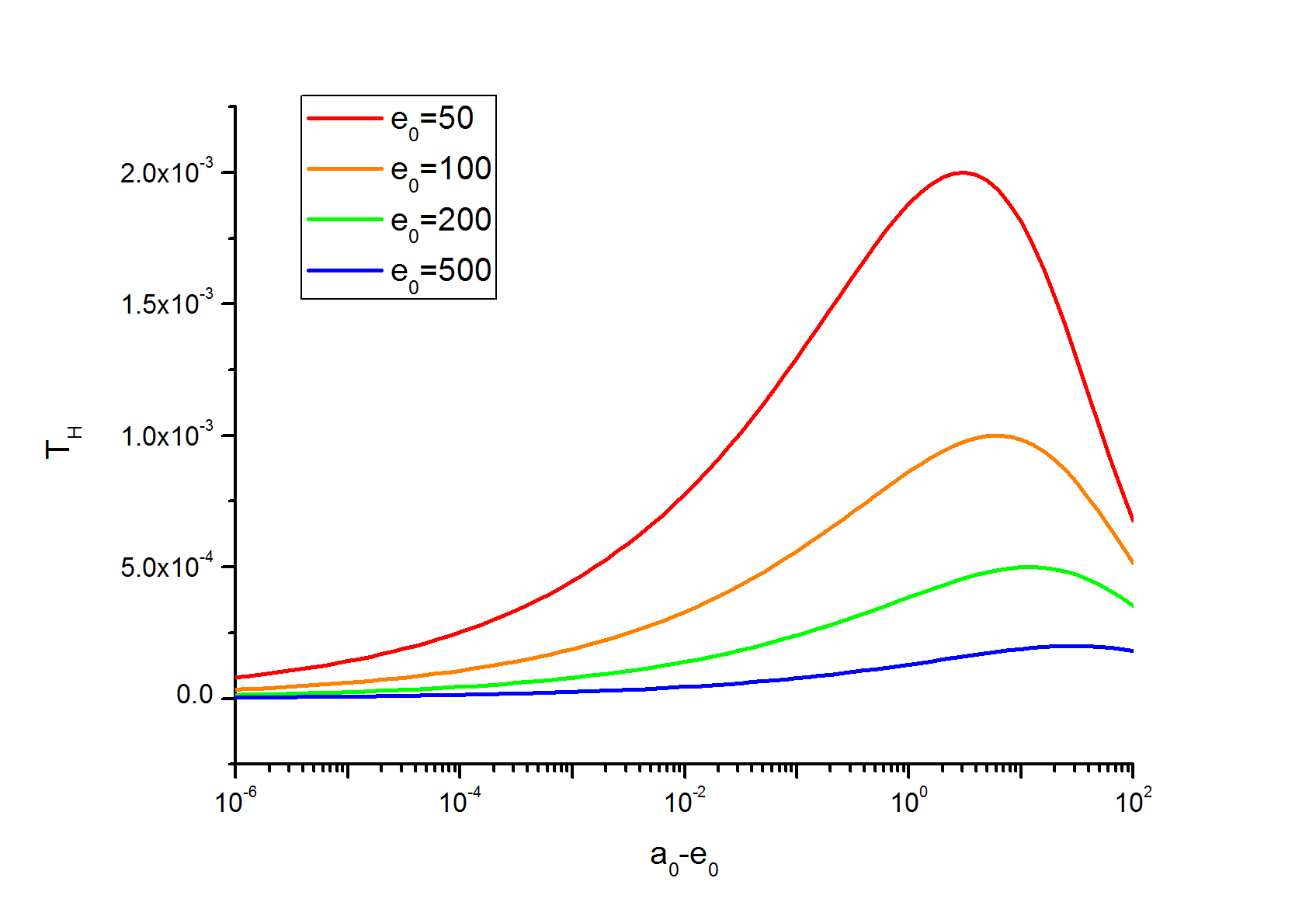}
  \caption{The plot of Hawking temperature against $a_{0}-e_{0}$. All the units are Planck units.\label{fig1}}
\end{figure}

On the other hand, the Hawking temperature approaches to 0 as the black hole becomes extremal; therefore, it reaches a maximum at some point of evaporation. (See figure \ref{fig1})

In order to find this temperature, one has to solve,
\begin{equation}
\partial_{a_{0}}J =0.
\end{equation}
The solution is 
\begin{equation}
a_{\textrm{ max}} =\frac{3e_{0}}{2\sqrt{2}}
\end{equation}
and,
\begin{equation}
T_{\textrm{max}} = \frac{(5/2)^{1/4}}{4\pi e_{0}}.
\end{equation}

Thus, for large $e_{0}$, the Hawking temperature will remain small for all periods of evaporation under the range $x \gg \ell_{p}^{4}/e_{0}^{3}$. In section \ref{sec:1-1}, We have stated that the approximation for chargeless radiation is reasonable because we assume that the Hawking radiation will always remain small during the entire evaporation process. We have showed that this is true if $x \gg \ell_{p}^{4}/e_{0}^{3}$, making this assumption consistent under this condition. 

Now we turn to the near-extremal case. Since there is little non-electromagnetic energy left, we can expect $\dot{a}_{0}$, and therefore, the Hawking temperature is very small.
To get a rough estimation, let $\dot{a}_{0} = \Delta a_{0} / \Delta u$. By taking $\Delta u = \ell_{p}$ and observing that $|\Delta a_{0}| < \ell_{p}^{4}/e_{0}^{3}$, we get $|\dot{a}_{0}| < \ell_{p}^{3}/e_{0}^{3}$, which is indeed very small.  

\subsection{The Energy Momentum Tensor}\label{sec:4-4}
\paragraph{ }
We shall calculate the energy momentum tensor $T_{\mu\nu}$ by inserting the RNV metric (\ref{eq:3-7}) into Einstein tensor.

The results are:
\begin{eqnarray}
   G_{uu} &=& \frac{e^{\psi}}{r^2}(r\partial_{u}F-e^{\psi}F(F-1+r\partial_{r}F)) \\
   G_{ur} &=& -\frac{e^{\psi}}{r^2}(F-1+r\partial_{r}F) \\
   G_{rr} &=& \frac{2\partial_{r}\psi}{r} \\
   G_{\theta\theta} &=& r(\partial_{r}F+F\partial_{r}\psi)+r^2\left(\frac{3}{2}\partial_{r}F\partial_{r}\psi+F(\partial_{r}\psi)^2
   +\frac{1}{2}\partial^{2}_{r}F+F\partial_{r}^2\psi-e^{-\psi}\partial_{r}\partial_{u}\psi\right) \\
   G_{\phi\phi} &=& \sin^2\theta G_{\theta\theta}
\end{eqnarray}
All the other components of Einstein tensor are zero.

Incorporating the condition of null infalling matter and obtaining from (\ref{eq:3-13})
\begin{equation}
\psi^{\prime} = 2\frac{ra^{\prime}-ee^{\prime}}{r^{2}F},
\end{equation}
and using the fact that the charge does not radiate out (i.e. $De = 0$), we can decompose the energy momentum potential into:
\begin{equation}
T_{\mu\nu}= T^{\textrm{(out)}}_{\mu\nu}+T^{\textrm{(in)}}_{\mu\nu}+T^{\textrm{(T)}}_{\mu\nu}+T^{\textrm{(EM)}}_{\mu\nu}.
\end{equation}
To express the energy momentum tensor, we adopt a set of null atlases being parallel to the one defined in (\ref{eq:1-2}),
\begin{eqnarray}
  l_{a} &=& e^{\psi}\delta_{a}^{0} \\
  n_{a} &=&  \frac{1}{2}Fe^{\psi} \delta_{a}^{0}+ \delta_{a}^{1} \\
  m_{a} &=& -\frac{r}{\sqrt{2}}\left( \delta_{a}^{2}+ i\sin\theta \delta_{a}^{3}\right)
\end{eqnarray}
where $T^{\textrm{(out)}}_{\mu\nu}$ represents the massless radiation:
\begin{equation}
T^{\textrm{(out)}}_{\mu\nu} = -\frac{Da}{4\pi G r^2}l_{\mu}l_{\nu},
\end{equation}
and $T^{\textrm{(in)}}_{\mu\nu}$ is the energy momentum tensor of the infalling matter:
\begin{equation}
T^{\textrm{(in)}}_{\mu\nu} = T_{\textrm{in}}n_{\mu}n_{\nu},
\end{equation}
where,
\begin{equation}
T_{\textrm{in}}=
\frac{a^{\prime}r-ee^{\prime}}{2G\pi r(r^{2}-2ar+e^{2})}.
\end{equation}
$T^{\textrm{(EM)}}_{\mu\nu}$ is the contribution of the electromagnetic field.

\begin{equation}
T^{\textrm{(EM)}}_{\mu\nu} = \rho n_{(\mu}l_{\nu)}+pm_{(\mu} \bar{m}_{\nu)}.
\end{equation}
where
\begin{equation}
\rho = p = \frac{e^2}{4\pi G r^4}.
\end{equation}
Finally, $T^{\textrm{(T)}}_{\mu\nu}$ is the tangential component induced by the collapse and the radiation.
\begin{equation}
T^{\textrm{(T)}}_{\mu\nu}=T_{\textrm{tan}}2m_{(\mu} \bar{m}_{\nu)}= T_{\textrm{tan}}r^2d\Omega.
\end{equation}

\begin{eqnarray*}
  T_{\textrm{tan}} &=& \frac{1}{8\pi r^2 G}\left(e^{\prime 2}-\frac{2}{r}ee^{\prime}+r\psi^{\prime}+a\psi^{\prime}-\frac{2e^{2}}{r}\psi^{\prime}-3ra\psi^{\prime}
  +3ee^{\prime}\psi^{\prime}\right.\\
   & &\left.+r^{2}F\psi^{\prime 2}+ra^{\prime\prime}+ee^{\prime\prime}
   +r^{2}F\psi^{\prime\prime}-e^{-\psi}\dot{\psi}^{\prime}\right).\\
\end{eqnarray*}

\section{Conclusion}
In this paper, we have proposed a model of the evaporation of charged black holes, which is an analog of KMY model proposed by Kawai  et al. We consider two scenarios: the thin-shell case and the generic spherical symmetric case.

In the thin-shell model, a charged sphere collapses at the speed of light. We demonstrated that an event horizon will form, and the shell will eventually fall into this horizon. Also, the black hole will not become extremal before it falls into the event horizon, provided that $a_{0}$ is not very close to $e_{0}$. That is, it does not radiate the maximal possible amount of energy before it falls into the event horizon.

We then considered a more realistic case: a continuous flow of charged matter that collapses spherically at the speed of light, which can be simulated by the continuous limit of many co-centered shells collapsing inward. The dynamics of the system is given at (\ref{eq:3-7}), (\ref{eq:3-8}), (\ref{eq:3-9}) and (\ref{eq:3-10}).

First, we considered a case that the black hole is not close to its extremal limit (in a sense that $(\ref{eq:2-6})$ holds), but has an extremal center covered by neutral outside.
The dynamics of the system is (\ref{eq:4-7}). The condition for this equation to hold is $a_{0} - e_{0} \gg \ell_{p}^{4}/e_{0}^{3}$. It is consistent with KMY model in the chargeless limit. ($e_{0} \ll a_{0}$) The Hawking temperature under this situation is depicted in figure \ref{fig1}. Unlike KMY model, the temperature is always very small for large $e_{0}$. More specifically, $T_{\textrm{max}} \sim 1/e_{0}$. This suggested that the field being chargeless is a reasonable assumption, since all charged fields have mass.  

We also considered the case that the black hole becomes near-extremal, in which the assumption for deriving (\ref{eq:4-7}) does not hold anymore. By only considering the null condition of each infalling shell, some statements about the the asymptotic behavior of $r_{\alpha}$ and $a_{\alpha}$ can be made. We classified different asymptotic behaviors by their limit values of $x_{\alpha}^{2}/y_{\alpha}$ (case \ref{case1}, case \ref{case2} and case \ref{case3}), and analyzed them separately. We found that, for case \ref{case1} and case \ref{case2}, $y_{\alpha} = A_{\alpha}/ u_{\alpha} + o(u_{\alpha}^{-1})$, and $x_{\alpha} = O(u_{\alpha}^{-2})$ (\ref{eq:4-23}-\ref{eq:4-25}), while for case \ref{case3}, there is a possibility that $x_{\alpha}$ will not approach to 0 in the late time limit.

Also, if we assume the weak energy condition, we can get an estimation of the redshift factor $\tilde{\psi}_{\alpha}$ in (\ref{eq:4-19}), and $\tilde{\psi}_{\alpha} \rightarrow -\infty$ in the late time limit. The general Penrose diagrams are also proposed in this section (Fig. \ref{fig:1-1} and Fig. \ref{fig:1-2}).

In conclusion, since in both cases (thin-shell and continuous), a permanent horizon forms, it implies that there will be no information loss problem, since the information can be kept inside the horizon.

However, there are still some questions that can be investigated in further researches. First, our analysis depended on a very specific regularization in calculating $\left\langle T_{\mu\nu}\right\rangle$, the point-splitting regularization, which was used to derive (\ref{eq:c}) in \cite{kawai2013self}. It is natural to ask how many of the important features (becoming extremal in late time limit, etc.) will remain if the regularization is changed to a different or more general regularization method. From the paper of Ho \cite{ho2015comment}, we know that the information will not be lost if $\dot{a}(u) < 0$. Therefore, it is also worth investigating whether there will be a negative energy flux for some regularization method if back reaction is considered (just as in KMY model). 

We can also generalize some of the assumptions in our analysis, such as the chargelessness of fields or the spherical symmetry. For example, we can consider a Kerr-like metric for a rotating black hole. The problem is that in our analysis, we simulated the infalling matter as a series of co-centered spheres, and by making Bogolubov transformation between the retarded times between connecting shells, one can get the flux equation $(\ref{eq:c})$. There is no simple generalization if the system is not spherical symmetric,and thus new ways of calculating the flux equation are needed.
\section{Acknoweldgement}
\paragraph{ }
The author would like to thank Pei-Ming Ho for his
guidance on the process of writing this paper. Also, he would like to thank Yi-Shih Chen and Yoshinori Matsuo for giving suggestions on this paper, and thank Pin-Chun Huang for drawing Fig.\ref{fig:1-2} and Fig. \ref{fig:1-1}.

\bibliographystyle{unsrt}
\bibliography{reference.bib}

\begin{thebibliography}{10}

\bibitem{hawking1975particle}
Stephen~W Hawking.
\newblock Particle creation by black holes.
\newblock {\em Communications in mathematical physics}, 43(3):199--220, 1975.

\bibitem{parker2009quantum}
Leonard Parker and David Toms.
\newblock {\em Quantum field theory in curved spacetime: quantized fields and
  gravity}.
\newblock Cambridge university press, 2009.

\bibitem{siahaan2010semiclassical}
Haryanto~M Siahaan and Triyanta.
\newblock Semiclassical methods for hawking radiation from a vaidya black hole.
\newblock {\em International Journal of Modern Physics A}, 25(01):145--153,
  2010.

\bibitem{triyanta2013hawking}
T~Triyanta and Anike~N Bowaire.
\newblock Hawking temperature of the reissner-nordstrom-vaidya black hole.
\newblock {\em Journal of Mathematical and Fundamental Sciences},
  45(2):114--123, 2013.

\bibitem{kawai2013self}
Hikaru Kawai, Yoshinori Matsuo, and Yuki Yokokura.
\newblock A self-consistent model of the black hole evaporation.
\newblock {\em International Journal of Modern Physics A}, 28(14):1350050,
  2013.

\bibitem{kawai2016interior}
Hikaru Kawai and Yuki Yokokura.
\newblock Interior of black holes and information recovery.
\newblock {\em Physical Review D}, 93(4):044011, 2016.

\bibitem{kawai2017model}
Hikaru Kawai and Yuki Yokokura.
\newblock A model of black hole evaporation and 4d weyl anomaly.
\newblock {\em Universe}, 3(2):51, 2017.

\bibitem{ho2015comment}
Pei-Ming Ho.
\newblock Comment on self-consistent model of black hole formation and
  evaporation.
\newblock {\em arXiv preprint arXiv:1505.02468}, 2015.

\bibitem{ho2016absence}
Pei-Ming Ho.
\newblock The absence of horizon in black-hole formation.
\newblock {\em Nuclear Physics B}, 909:394--417, 2016.

\bibitem{vaidya1951gravitational}
Prahalad~Chunnilal Vaidya.
\newblock The gravitational field of a radiating star.
\newblock {\em Proceedings Mathematical Sciences}, 33(5):264--276, 1951.

\bibitem{boulware1976hawking}
David~G Boulware.
\newblock Hawking radiation and thin shells.
\newblock {\em Physical Review D}, 13(8):2169, 1976.

\bibitem{ibohal2002variably}
Ng~Ibohal.
\newblock On the variably-charged black holes in general relativity: Hawking's
  radiation and naked singularities.
\newblock {\em Classical and Quantum Gravity}, 19(16):4327, 2002.

\bibitem{ibohal2005rotating}
Ng~Ibohal.
\newblock Rotating metrics admitting non-perfect fluids.
\newblock {\em General Relativity and Gravitation}, 37(1):19--51, 2005.

\bibitem{ibohal2010charged}
Ng~Ibohal and L~Kapil.
\newblock Charged black holes in vaidya backgrounds: Hawking's radiation.
\newblock {\em International Journal of Modern Physics D}, 19(04):437--464,
  2010.

\bibitem{wang1999generalized}
Anzhong Wang and Yumei Wu.
\newblock Generalized vaidya solutions.
\newblock {\em General Relativity and Gravitation}, 31(1):107--114, 1999.

\bibitem{ho2016asymptotic}
Pei-Ming Ho.
\newblock Asymptotic black holes.
\newblock {\em arXiv preprint arXiv:1609.05775}, 2016.

\end{thebibliography}

\end{document}